\documentclass{article}
\usepackage{spconf,amsmath,graphicx}
\usepackage{url}
\usepackage{subfig}
\usepackage{xcolor}
\usepackage{cite}
\usepackage{hyperref}
\usepackage{cleveref}
\usepackage{booktabs, multirow} 
\usepackage{soul}
\usepackage{amsfonts}
\usepackage{bm}

\title{VISinger2+: End-to-End Singing Voice Synthesis Augmented by Self-Supervised Learning Representation}
\name{Yifeng Yu$^{\star}$, Jiatong Shi$^{\dagger}$, Yuning Wu$^{\ddagger}$, Yuxun Tang$^{\ddagger}$, Shinji Watanabe$^{\dagger}$}

\address{
    $^{\star}$ Georgia Institute of Technology \\
    $^{\dagger}$ Carnegie Mellon University \\
    $^{\ddagger}$ Renmin University of China \\
}

\begin{document}
%
\maketitle

\begin{abstract}
Singing Voice Synthesis (SVS) has witnessed significant advancements with the advent of deep learning techniques. However, a significant challenge in SVS is the scarcity of labeled singing voice data, which limits the effectiveness of supervised learning methods. In response to this challenge, this paper introduces a novel approach to enhance the quality of SVS by leveraging unlabeled data from pre-trained self-supervised learning models. Building upon the existing VISinger2 framework, this study integrates additional spectral feature information into the system to enhance its performance. The integration aims to harness the rich acoustic features from the pre-trained models, thereby enriching the synthesis and yielding a more natural and expressive singing voice. Experimental results in various corpora demonstrate the efficacy of this approach in improving the overall quality of synthesized singing voices in both objective and subjective metrics.
\end{abstract}

\begin{keywords}
Singing voice synthesis, self-supervised learning
\end{keywords}

\section{Introduction}
\label{sec: intro}

Singing voice synthesis (SVS) is a captivating field that aims to create realistic and expressive singing voices from music scores and lyrics. Its applications span across music production and entertainment, revolutionizing how music is created and experienced. Traditional SVS methodologies, exemplified by VOCALOID~\cite{KenmochiO07} and UTAU~\cite{ameya_ayame}, have long dominated the field, utilizing concatenative synthesis to stitch together pre-recorded vocal samples into a coherent singing voice~\cite{bonada2001singing, bonada2003sample}. Despite their pioneering role, these systems often necessitate extensive manual adjustments to produce satisfactory outcomes. In contrast, deep learning has ushered in a new paradigm of SVS systems ~\cite{lu2020xiaoicesing, chen2020hifisinger, VISinger, visinger2, liu2022diffsinger}, such as \href{https://ace-studio.timedomain.cn/}{ACE Studio} and \href{https://dreamtonics.com/synthesizerv/}{Synthesizer V}, which leverage neural networks to emulate complex musical expressions and dynamics. These modern engines, trained on expansive datasets, are capable of delivering a diverse array of vocal styles, thus offering unparalleled customization and versatility in music production. This paradigm shift not only enhances the quality of synthesized singing but also simplifies the creation process, marking a significant milestone in the ongoing evolution of SVS.

Building on this foundation, the SVS through artificial approaches has seen remarkable advancements in recent years, driven by significant breakthroughs in deep learning. These advancements include developments in non-autoregressive models \cite{lu2020xiaoicesing, shi2021sequence, chunhui23_interspeech, chen2020hifisinger}, diffusion models \cite{liu2022diffsinger}, and end-to-end models \cite{VISinger, visinger2, wu2023systematic}. Among the various systems developed for this purpose, VISinger \cite{VISinger} has emerged as a prominent framework for generating high-quality singing voices. Building on this success, VISinger2 \cite{visinger2} has been introduced as an improved version of VISinger, further enhancing the audio quality.

Despite its successes, there remains a continuous quest for improvement, particularly in enhancing the richness and expressiveness of the synthesized voice. A primary challenge is the scarcity of data. The high cost and complexity of acquiring and annotating singing voice data make it difficult to obtain large-scale, high-quality labeled datasets. This data scarcity issue limits the potential for training robust and expressive SVS models~\cite{shi2021sequence, guo2022singaug, shi2024singing}.

One of the promising solutions to the data scarcity issue is to utilize unlabeled data. Specifically, pre-trained self-supervised learning~(SSL) models have revolutionized the field of audio and speech processing by extracting better representation without explicit labeling~\cite{hubert, yizhi2023mert, baevski2022data2vec, shi2024multiresolution}. It has been known that SSL-based methods have their superiority in speech and music understanding tasks~\cite{yang21c_interspeech, mlsuperb, yuan2024marble}. While recent studies also reveal that the integration of SSL models has shown promising results in enhancing the quality of generated speech and audio, suggesting their applicability to SVS ~\cite{kakoulidis2024low, fujita2024noise, wang23_ssw, sivaguru23_interspeech, Shah2024, yang2023towards, gong2023zmm}.

This study introduces a new framework, namely VISinger2+, that leverages the strengths of pre-trained SSL models (e.g., HuBERT~\cite{hubert} and MERT~\cite{yizhi2023mert}) to enrich the spectral inputs used in VISinger2~\cite{visinger2}. We conduct comprehensive experiments with our methods in both single-singer and multi-singer scenarios, as well as in different languages, such as Japanese and Mandarin. Our findings show that the proposed VISinger2+ can produce better quality singing voices than the baseline VISinger2 in most subjective and objective metrics. Our open-source implementation can be found at \footnote{\url{https://espnet.github.io/espnet/recipe/svs1.html\#visinger-2-plus-training}}.
\section{Method}
\label{sec: method}

VISinger2+ introduces a novel enhancement to the original VISinger2 architecture through the integration of pre-trained SSL models. This section delineates the modified architecture and the workflow employed to synthesize singing voices.

\subsection{Overview of VISinger2 Framework}

\begin{figure}[t]
  \centering
  \includegraphics[width=0.5\textwidth]{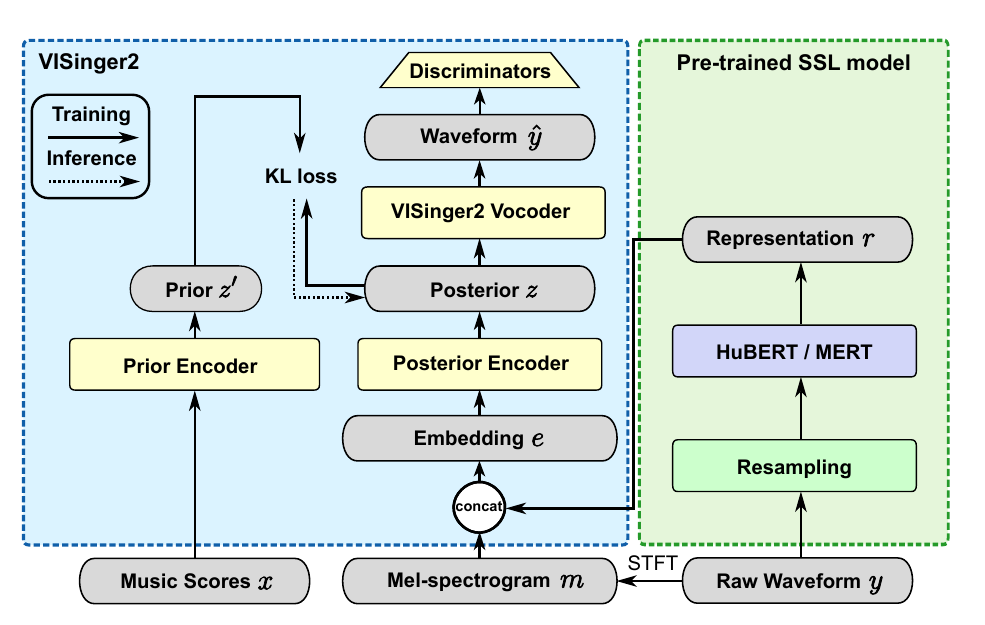}
  \caption{The proposed VISinger2+ architecture. Model details are discussed in Section~\ref{sec: method}.}
  \label{fig:model}
\end{figure}


The original VISinger2 framework \cite{visinger2}, as shown in the left side of Figure \ref{fig:model}, composed of a prior encoder, a posterior encoder, a vocoder, and discriminators, forms the core of our SVS system. In this framework, let \( \bm{x} \) denote the input music scores, which include pitch, phoneme, and duration information, and let \( \bm{y} \) represent the waveform of the singing voice, which is the target output of our synthesis system.

The prior encoder, the posterior encoder, and the vocoder are jointly utilized to generate and process the latent representation of the singing voice:

\begin{itemize}
    \item The \textbf{prior encoder} processes the input music scores \( \bm{x} \) to generate a prior distribution \( \bm{z'} \) over the latent space. This distribution encapsulates the expected features of the singing voice based on the musical context:
    \begin{equation}
        \bm{z'} = \text{Enc}_{\text{pri}}(\bm{x}) \sim q_{\text{pri}}(\bm{z'}|\bm{x})
    \end{equation}
    
    \item The \textbf{posterior encoder} takes the waveform \( \bm{y} \) as input and maps it to a latent representation \( \bm{z} \) in the same latent space in the training phase. This representation is assumed to follow the posterior distribution \( q_{\text{post}}(\bm{z}|\bm{m}) \), capturing the actual features of the singing voice:
    \begin{equation}
        \bm{z} = \text{Enc}_{\text{post}}(\bm{y}) \sim q_{\text{post}}(\bm{z}|\bm{m})
        \label{eq:posterior_encoder}
    \end{equation}

    It is worth noting that in the original VISinger2, the Mel-spectrogram \( \bm{m} \) is directly inputted into the posterior encoder, while the proposed method incorporates additional embedding \( \bm{e} \), as shown in Figure \ref{fig:model}, which will be introduced in Section \ref{subsec: integration}.

    \item The \textbf{vocoder} is crucial in both the training and inference phases of the VISinger2 framework:
    
    \begin{itemize}
        \item \textbf{Training Phase}: During training, the vocoder is optimized to reconstruct the waveform \(\bm{y}\) based on the latent representation \(\bm{z}\) obtained from the posterior encoder:
        \begin{equation}
            \hat{\bm{y}} = \text{Voc}(\bm{z}) \sim p(\bm{y}|\bm{z})
        \end{equation}
        Here, \(\hat{\bm{y}}\) represents the vocoder's reconstruction of the original singing voice waveform \(\bm{y}\). The reconstruction is drawn from the distribution \(p(\bm{y}|\bm{z})\), which represents the likelihood of the singing voice given the latent representation \(\bm{z}\) from the posterior encoder's output during training.
    
        \item \textbf{Inference Phase}: During inference, the vocoder generates the synthesized singing voice based on the latent representation \(\bm{z'}\) obtained from the prior encoder:
        \begin{equation}
            \hat{\bm{y}} = \text{Voc}(\bm{z'}) \sim p(\bm{y}|\bm{z'})
        \end{equation}
        In this phase, the synthesized singing voice \(\hat{\bm{y}}\) is produced by modeling the conditional distribution \(p(\bm{y}|\bm{z'})\), where \(\bm{z'}\) captures the expected features of the singing voice based on the input music scores from the prior encoder.
    \end{itemize}
\end{itemize}

\subsection{Integration with pre-trained SSL Models}
\label{subsec: integration}


In our enhanced system VISinger2+, we leverage the pre-trained SSL models, as depicted on the right side of Figure \ref{fig:model}, to enrich the feature set provided to the posterior encoder.

\begin{itemize}
    \item \textbf{Resampling}: Given the difference in sampling rates and feature scales between the VISinger2 model and the pre-trained SSL models, a resampling adjustment is required. this adjustment is achieved through a DSP-based resampling method. Specifically, the sample rate of the waveform \( \bm{y} \) is adjusted to be compatible with the sample rate of the pre-trained SSL model. The resampling is performed using the sinc interpolation with a Hann window method.

    \item \textbf{Feature Extraction from the Pre-trained Model}: The resampled waveform is fed into the pre-trained SSL model to extract multi-layer SSL representations. In previous SSL-related works, representations from different layers were shown to exhibit different information~\cite{pasad2021layer, chen2022wavlm, mlsuperb, shi2024multiresolution}. To efficiently utilize SSL representations for SVS, we adopt a weighted-sum strategy, following SUPERB~\cite{yang21c_interspeech},  to aggregate multiple hidden states from the pre-trained SSL model. Let \( \bm{h}_i \) be the hidden state at the \( i \)-th layer of the model, and \( \alpha_i \) be the learnable weight corresponding to this layer. The final representation \( \bm{r} \) is obtained as follows:

    \begin{equation}
        \bm{r} = \sum_{i=1}^{L} \alpha_i \bm{h}_i,
    \end{equation}
    where \( L \) is the total number of layers from which hidden states are considered. This strategy moves beyond using just the last-layer representation, forming a robust and comprehensive final representation of the audio input.


    \item \textbf{Feature Integration}: The representation \( \bm{r} \), with dimensions \(( B, F, 1024 )\), is concatenated along the feature dimension with the original Mel-spectrogram \(\bm{m}\), which has dimensions \(( B, F, 80 )\). This concatenation results in the embedding \(\bm{e}\), which has dimensions \(( B, F, 1104 )\), effectively enriching the feature set for the subsequent processing steps. Here, \(B\) represents the batch size, and \(F\) represents the frame size. 
    This concatenation integration method is more straightforward and does not require the introduction of additional modules, which would increase complexity.

    \begin{equation}
        \bm{e} = \text{Concat}(\bm{r}, \bm{m})
    \end{equation}

    In the proposed VISinger2+ framework, the input of the posterior encoder is modified to take the embedding \(\bm{e}\) as input instead of the Mel-spectrogram \(\bm{m}\), as shown in the previous Equation \ref{eq:posterior_encoder}:
    \begin{equation}
        \bm{z} = \text{Enc}_{\text{post}}(\bm{y}) \sim q_{\text{post}}(\bm{z}|\bm{e})
    \end{equation}
    
\end{itemize}

\subsection{Training and Loss Function}

The entire system is trained end-to-end. The training objective includes minimizing the KL loss between the prior \(\bm{z'}\) and posterior \(\bm{z}\) distributions, which is crucial for the accurate generation of singing voices that are consistent with the input music scores. The discriminators including Multi-Resolution Spectrogram Discriminator (MRSD) \cite{jang21_interspeech}, Multi-Period Discriminator (MPD), and Multi-Scale Discriminator (MSD) \cite{kong2020hifi}, are identical to those used in the original VISinger2 framework. These discriminators employ adversarial loss to refine the Vocoder's output, ensuring the synthesis of high-fidelity waveforms. The Discriminator and loss functions are consistent with those delineated in the original VISinger2 paper.

\section{Experiments}

\subsection{Experimental Datasets}

This study utilizes three datasets in Mandarin and Japanese to evaluate the proposed framework. The selected datasets offer a diverse range of singing styles, from single-singer to multi-singer, providing a comprehensive basis for assessing the effectiveness of the implemented techniques.

The Opencpop \cite{wang22b_interspeech} dataset, a publicly available high-quality Mandarin singing corpus, is specifically designed for SVS systems. It includes 100 unique Mandarin songs, recorded by a professional female single-singer, amounting to approximately 5.2 hours of audio. These songs were recorded in a studio environment with a sampling rate of 44,100 Hz and meticulously annotated for utterance, note, phoneme boundaries, and pitch types. This dataset contains 3,756 utterances.

The Ofuton-P \cite{Ofuton} dataset is a Japanese singing voice database with 56 songs, totaling 61 minutes of audio, recorded by a single male vocalist. It offers a distinctive perspective on Japanese singing styles, crucial for testing the adaptability of SVS systems in different linguistic and vocal contexts.

Additionally, the ACE-Opencpop \cite{shi2024singing} dataset, a multi-singer Mandarin dataset generated by ACE Studio, contains more than 100 hours of singing voices. Contrasting with the Opencpop and Ofuton-P datasets, which focus on single-singer SVS, the ACE-Opencpop dataset comprises a variety of vocal characteristics and styles from 30 singers. This diversity is vital for evaluating the robustness and generalizability of the proposed SVS system enhancements.

For these three corpora, we followed the recipe from Muskits-ESPnet \cite{shi2022muskits} \cite{10.1145/3664647.3685000} to split the data into training, validation, and test sets.

\subsection{Experimental Models}
The baseline in our experiments is the original VISinger2~\cite{visinger2} trained on three corpora, respectively. For our proposed VISinger2+, we have selected three candidate SSL models for fusion: the official HuBERT-large~\cite{hubert}, MERT-large-300M~\cite{yizhi2023mert}, and Chinese HuBERT-large (CN-HuBERT)\footnote{\scriptsize{\url{https://huggingface.co/TencentGameMate/chinese-hubert-large}}} available on Huggingface. These models are chosen for their distinct characteristics, particularly in terms of pre-trained languages and domains. HuBERT-large is pre-trained on English speech, MERT-large-300M is tailored for music data, and CN-HuBERT is specialized in Mandarin speech. By evaluating these models, we aim to provide deeper insights into how the pre-training languages and domains contribute to the VISinger2+ framework. It should be noted that we do not include Ofuton-P in our experiments with CN-HuBERT, given that language inconsistency has been sufficiently assessed in experiments based on HuBERT.

\subsection{Experimental Setups}


\noindent \textbf{Audio Processing Parameters}:
    In our experimental setup, the audio was processed at a sampling rate of 24,000 Hz with a hop size of 480 for the Mel-spectrogram features. These settings were specifically chosen to align the frame size of \( \bm{m} \) with \( \bm{r} \) in Section \ref{subsec: integration}, ensuring consistent feature dimensions. Both the number of FFT points and the window size were set to 2048, with a Hann window function applied. 80 Mel basis filters were used, spanning a frequency range from 0 Hz to 12,000 Hz. For the HuBERT and CN-HuBERT models, the spectral inputs were resampled to 16,000 Hz to align with the SSL models' input requirements as mentioned in Section \ref{subsec: integration}. For MERT, the input sample rate was already at 24,000 Hz and therefore required no changes.
    
\noindent  \textbf{Training Configuration}:
    The training of the VISinger2+ model was conducted over 200 epochs, with each epoch consisting of 1000 iterations. The optimization was carried out using the AdamW optimizer, configured with a learning rate of \(2.0 \times 10^{-4}\), betas set to [0.8, 0.99], an epsilon of \(1.0 \times 10^{-9}\), and no weight decay. The learning rate was scheduled to decrease exponentially with a gamma value of 0.998. This training setup was the same as the original VISinger2 model. In our implementation, the decoder channels were set to 512. To accommodate multi-singer settings, we assign unique speaker labels as additional embeddings for the model, allowing it to learn and adapt to the vocal characteristics of each singer.

\subsection{Evaluation}

The evaluation of the proposed VISinger2+ model was conducted using a comprehensive set of objective and subjective measures to assess the quality of the synthesized singing voice. The primary experiments were conducted on the Opencpop dataset, while the ablation studies were carried out on the Ofuton-P and ACE-Opencpop datasets. Therefore, we only conducted the subjective evaluation on the Opencpop dataset.

\begin{itemize}

\item \textbf{Objective Evaluation}:
For objective evaluation,  following previous works~\cite{shi2022muskits, guo2022singaug, huang2023singing}, several standard metrics were employed to quantify the performance of the SVS system. These metrics included Mean Mel-Cepstral Distortion (MCD), Root Mean Square Error of logarithmic fundamental frequency (F0 RMSE), and semitone accuracy (ST Acc).

We use SingMOS~\cite{tang2024singmos} as an additional MOS evaluation for Ofuton-P and ACE-Opencpop datasets, which lack human MOS evaluations. SingMOS leverages a pre-trained MOS prediction model to generate MOS-annotated results for singing voices. This approach enables efficient and reliable subjective ratings for these datasets.

Additionally, for the multi-singer ACE-Opencpop dataset, Speaker Embedding Cosine Similarity (SECS) was utilized to evaluate the similarity between different singers. SECS measures the closeness of the synthesized voice to the target singer's voice, thus assessing the model's ability to capture individual singer characteristics. We used the Rawnet-based speaker embedding extractor~\cite{jung20c_interspeech} pre-trained in ESPnet-SPK~\cite{jung2024espnet}.

\item \textbf{Subjective Evaluation}:
Subjective evaluation was conducted using the Mean Opinion Score (MOS) test. In this test, listeners were asked to rate a random set of 30 groups of samples, each with a 1-5 integer scoring system. Each group of samples consisted of five versions: one synthesized using the HuBERT model as additional information, one with CN-HuBERT, one with MERT, one baseline (VISinger2), and the ground truth recording, totaling 30 samples per group. For this test, we selected 20 individuals to perform the ratings.
\end{itemize}
\section{Results}

This section presents the experimental results obtained from evaluating the VISinger2+ framework across three distinct datasets: Opencpop, Ofuton-P, and ACE-Opencpop. The performance of the models is evaluated using both objective MCD and F0 RMSE and the subjective MOS, where applicable.

\begin{table}[t]
  \caption{Experimental results on the Opencpop dataset in three objective metrics and subjective mean opinion score (MOS).}
  \label{tab:opencpop}
  \centering
  \resizebox{\linewidth}{!}{%
  \begin{tabular}{l|ccccc}
    \toprule
    \textbf{Model} & \textbf{MCD $\downarrow$} & \textbf{F0 RMSE $\downarrow$} & \textbf{ST Acc $\uparrow$} & \textbf{MOS $\uparrow$} \\
    \midrule
    VISinger2 & 7.625 & 0.177 & 60.55\% & 3.65 ($\pm$ 0.05) \\
    \midrule
    VISinger2+ (HuBERT) & 7.647 & 0.171 & 62.78\% & 3.69 ($\pm$ 0.05) \\
    VISinger2+ (CN-HuBERT) & \textbf{7.518} & 0.174 & 62.80\% & 3.71 ($\pm$ 0.05) \\
    VISinger2+ (MERT) & 7.602 & \textbf{0.167} & \textbf{62.97\%} &  \textbf{3.72 ($\pm$ 0.05)} \\
    \midrule
    G.T. & - & - & - & 4.57 ($\pm$ 0.05) \\
    \bottomrule
  \end{tabular}%
  }
\end{table}

\begin{table}[t!]
  \caption{Experimental results on the Ofuton-P dataset with objective evaluation.}
  \label{tab:ofuton}
  \centering
  \resizebox{\linewidth}{!}{%
  \begin{tabular}{l|ccccc}
    \toprule
    \textbf{Model}  & \textbf{MCD $\downarrow$} & \textbf{F0 RMSE $\downarrow$} & \textbf{ST Acc $\uparrow$}  & \textbf{SingMOS $\uparrow$} \\
    \midrule
    VISinger2 & 5.735 & \textbf{0.088} & 65.94\% & 3.42 ($\pm$ 0.08)  \\
    \midrule
    VISinger2+ (HuBERT) & \textbf{5.687} & 0.092 & \textbf{66.40\%} & 3.48 ($\pm$ 0.08)   \\
    VISinger2+ (MERT) & 5.750 & \textbf{0.088} & 66.15\% & \textbf{3.49 ($\pm$ 0.09)} \\
    \midrule
    G.T. & - & - & - & 3.53 ($\pm$ 0.10) \\
    \bottomrule
  \end{tabular}%
  }
\end{table}

\begin{table*}[t!]
  \caption{Experimental results on the ACE-Opencpop dataset with objective evaluation.}
  \label{tab:ace}
  \centering
  \begin{tabular}{l|ccccc}
    \toprule
    \textbf{Model}  & \textbf{MCD $\downarrow$} & \textbf{F0 RMSE $\downarrow$} & \textbf{ST Acc $\uparrow$} & \textbf{SECS $\uparrow$}   & \textbf{SingMOS$\uparrow$}  \\
    \midrule
    VISinger2 & 5.399 & \textbf{0.135} & 63.62\% & 0.795 & 3.74 ($\pm$ 0.01) \\
    \midrule
    VISinger2+ (HuBERT) & 5.234 & 0.140 & \textbf{65.36}\% & 0.833 & 3.82 ($\pm$ 0.01) \\
    VISinger2+ (CN-HuBERT) & 5.259 & 0.145 & 65.02\% & \textbf{0.836} & 3.78 ($\pm$ 0.01) \\
    VISinger2+ (MERT) & \textbf{5.181} & 0.141 & 64.93\% &  0.829 & \textbf{3.85 ($\pm$ 0.01)} \\
    \midrule
    G.T. & - & - & - & - & 3.85 ($\pm$ 0.01) \\
    \bottomrule
  \end{tabular}%
\end{table*}

\begin{figure*}[t!]
  \centering
  \includegraphics[width=\textwidth]{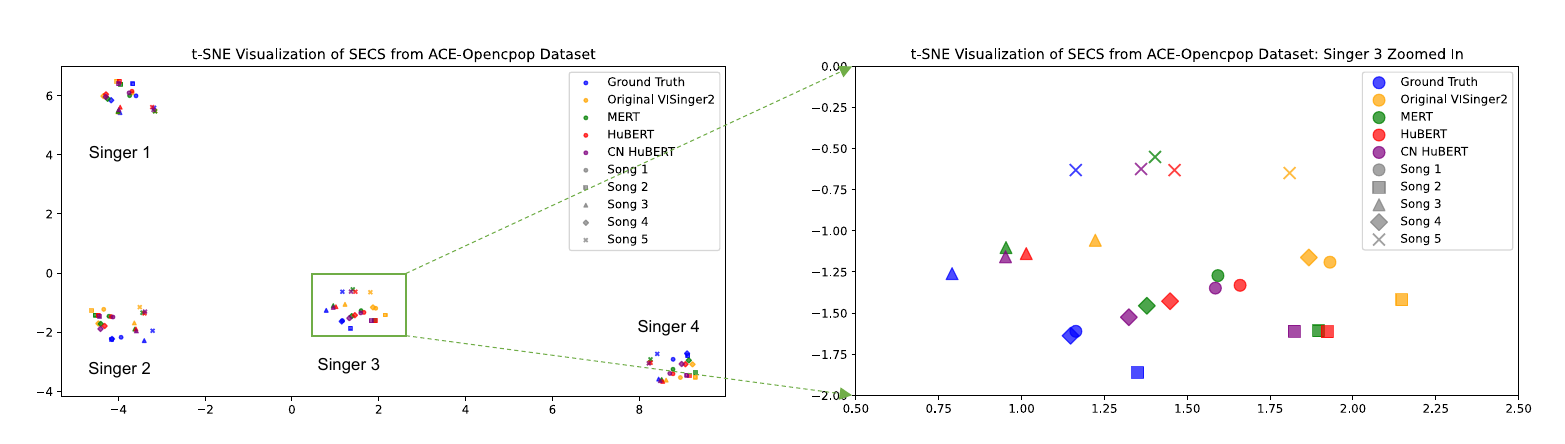}
  \caption{t-SNE Visualization of singer embeddings on synthesized audios from 4 singers in the ACE-Opencpop dataset.}
  \label{fig:singer_emb}
\end{figure*}

Across the datasets in Table \ref{tab:opencpop}, \ref{tab:ofuton}, and \ref{tab:ace}, VISinger2+ consistently improved the MCD values compared to the original VISinger2, indicating a closer spectral similarity to the ground truth singing voices. This improvement suggests that the feature representations learned by HuBERT and MERT are effective in enhancing the spectral quality of the synthesized singing voices. Notably, the VISinger2+ (CN-HuBERT) exhibited the most significant improvement in MCD on the Opencpop dataset, while the integration of MERT showed superior performance on the ACE-Opencpop dataset. These variations highlight the strengths of each pre-trained model in capturing different aspects of audio information that contribute to the overall quality of singing voice synthesis.

The impact on F0 RMSE was more variable, with some configurations showing a slight increase in error especially in the ACE-Opencpop dataset. This variability indicates the complex balance between achieving spectral fidelity and maintaining accurate pitch representation in the synthesized voices. Despite these variations, the enhancements in predicted spectral features, as evidenced by the lower MCD values, suggest that the trade-offs may be favorable for overall sound quality. In contrast to the variability observed in F0 RMSE, the Semitone Accuracy (ST Acc) consistently improved across all configurations. This implies that while there may be some deviations in pitch details, there is an overall enhancement at a macro level.

The SingMOS scores provide further insights into the perceptual quality of the synthesized singing voices. As seen in Table \ref{tab:ofuton}, the VISinger2+ (MERT) model achieved the highest SingMOS score of 3.49 on the Ofuton-P dataset, closely matching the ground truth score of 3.53. Similarly, Table \ref{tab:ace} shows that the VISinger2+ (MERT) model also obtained the highest SingMOS score of 3.85 on the ACE-Opencpop dataset, equaling the ground truth score. These results indicate that the VISinger2+ models not only enhance objective metrics but also significantly improve the perceived quality of the synthesized singing voices.

Furthermore, the integration of the pre-trained SSL models showed promising results in improving singer-similarity for multi-singer models. As shown in Table \ref{tab:ace}, VISinger2+ (CN-HuBERT) achieves the highest singer similarity. Furthermore, all VISinger2+ models utilizing various pre-trained SSL models surpass the performance of the original VISinger2. We also visualize the SECS results by plotting singer embeddings, as illustrated in Figure \ref{fig:singer_emb}. We selected SVS audio samples of four singers, each with five songs using the singer embeddings extracted by the pre-trained models in ESPnet-SPK~\cite{jung2024espnet}.\footnote{\scriptsize{\url{https://huggingface.co/espnet/voxcelebs12_rawnet3}}} For each song, we computed the average embeddings across 36 utterances to obtain five average embeddings per song. The left side of the figure shows the overall proximity of the embeddings to the ground truth, while the right side provides a detailed view of one selected singer. It can be observed that for each of the five songs, the embeddings from the proposed VISinger2+ are consistently closer to the ground truth compared to those from the original VISinger2. This indicates that our proposed VISinger2+ not only enhances the spectral quality of the synthesized singing voices but also contributes to a more accurate representation of the singer's identity. This improvement in singer-similarity is particularly beneficial for multi-singer synthesis systems to ensure each voice remains unique and realistic.

The subjective evaluation of the Opencpop dataset, as shown in Table \ref{tab:opencpop}, reveals the effectiveness of the VISinger2+ framework in enhancing the perceptual quality of SVS. The MOS results indicate that all variants of VISinger2+ outperform the original VISinger2, with VISinger2+ (MERT) achieving the highest score of 3.72. The confidence interval ($\pm$ 0.05) suggests that the perceived quality differences between the models are statistically significant, further validating the improvements brought by the integration of the pre-trained SSL models.

\section{Conclusion}\label{sec:concl}
In this paper, we introduced VISinger2+, an enhanced VAE-based framework for SVS that leverages pre-trained SSL models to enrich the input feature set beyond the traditional Mel-spectrogram. Our experiments across multiple datasets demonstrated that VISinger2+ consistently outperforms the original VISinger2 in terms of spectral similarity and singer-similarity. The integration of SSL models such as HuBERT, CN-HuBERT, and MERT contributed to a noticeable improvement in the overall quality of the synthesized singing voices. In future work, we intend to refine our approach by fine-tuning the pre-trained models for singing voice characteristics and exploring the fusion of diverse features to enhance synthesis performance. To ensure the reproducibility of our research, we will make the source code and model checkpoints publicly available upon the publication of this paper.



\bibliographystyle{IEEEbib}
\bibliography{strings,refs}

\end{document}